# Are We Approaching the Fundamental Limits of Wireless Network Densification?

Jeffrey G. Andrews, Xinchen Zhang, Gregory D. Durgin, Abhishek K. Gupta[1]

May 28, 16

## Abstract

The single most important factor enabling the data rate increases in wireless networks over the past few decades has been *densification*, namely adding more base stations and access points and thus getting more spatial reuse of the spectrum. This trend is set to continue into 5G and beyond. However, at some point further densification will no longer be able to provide exponentially increasing data rates. Like the end of Moore's Law, this will have extensive implications on the entire technology landscape, which depends ever more heavily on wireless connectivity. When and why will this happen? How might we prolong this from occurring for as long as possible? These are the questions explored in this paper.

## 1    The Importance of Densification

Wireless communication (per link) data rates have doubled about every two years since the advent of the cellular phone, an observation sometimes referred to as "Cooper's Law", after cell phone pioneer Martin Cooper. Over the last ten years, corresponding roughly with the advent of the "smart phone", that pace has significantly quickened to as fast as an annual doubling, as well-documented by the informative Cisco Visual Networking Index annual reports, which continues to forecast unrelenting growth through 2020.

This phenomenal long-term growth – spanning over 3 decades – has been primarily accommodated by network densification and the technological advances necessary to support such densification. By *densification*, we mean the deployment of more base stations and access points per unit area (or volume). Although an increase in the amount of spectrum and the spectral efficiency have also played a role in Cooper's Law, these two mechanisms combined have historically been a fairly small fraction of the total growth. For example, the combined amount of spectrum available for licensed mobile broadband use (namely cellular systems) and unlicensed wireless local area networking has barely increased since the turn of the millennium, and is roughly 500 MHz each. Although LTE's spectral efficiency does exceed that of previous 3GPP standards (Release 7 and prior), the increase is much less than widely believed, on the order of a factor of 2 versus HSPA according to a variety of industry and 3GPP documents. Similarly, WiFi spectral efficiencies have only incrementally improved from 802.11a which was released in 1999: other than the introduction of spatial multiplexing, most of the claimed data rate increases in subsequent releases are due to channel bonding (using more bandwidth) and optional modes that are seldom viable like 1024 QAM or multiuser MIMO.

---

[1] J. Andrews, X. Zhang, and A. Gupta are with the University of Texas at Austin.
G. Durgin is with the Georgia Institute of Technology.
The contact author is J. Andrews, jandrews@ece.utexas.edu.

As far as increasing data rates through increased spectrum, the scope for "beachfront spectrum" – namely licensed spectrum below 5 GHz with broad geographic support – is very limited. As a result, several new paradigms for dynamic spectrum access (DSA) are being explored. For example, in the United States there is a 100 MHz band at 3.5 GHz band that is currently used by the US Department of Defense for radar, mainly by about a dozen aircraft carriers worldwide. This band is likely to be made available for use subject to the restriction that the band is vacated when there is an aircraft carrier in the vicinity; network operators would only have a 60 second warning before having to vacate all active users from that spectrum. That such restrictions are being entertained as reasonable speaks to how scarce good broadband spectrum is. Meanwhile, spectrum approaching millimeter wavelengths (mmWaves, e.g. above 15 GHz) is certainly more plentiful, but is still at best unproven for cellular and WiFi use, since such frequencies struggle to penetrate most walls or to support significant mobility given the requisite beamforming. Furthermore, it is widely believed that mmWave networks will be extremely dense in order to overcome their propagation challenges. Although making better use of spectrum is certainly important and will delay any negative effects of ultra-densification, the results discussed in this paper apply equally (and possibly even more dramatically) to systems using DSA or mmWave.

## 2   The End of Cooper's Law?

Therefore, the data rate gains demanded by many emerging applications will require significant further network densification. Qualcomm, the world's leading cellular chipmaker, publicly states that "more small cells is the foundation of 1000x'"– a 1000x increase in mobile data network throughput – and further claims that capacity scales linearly with the number of small cells added to a network. Thus, according to them, doubling the number of base stations doubles the capacity; keep doing that indefinitely, problem solved.

But what if network densification stopped delivering significant throughput gains? Like all exponential trends, "Cooper's Law" must eventually hit a plateau. The obvious analogy is to the better known Moore's Law, which has recently hit such a plateau, resulting in wide ranging ramifications beyond the semiconductor industry. No longer can we expect chips to be ever smaller and lower power, and computers to be ever faster (at the same power and size). The end of Moore's Law poses major challenges to many sectors of the high-tech ecosystem.

In a decade, wireless communication technologies will be nearly as ubiquitous as chips, connecting not only people and their personal devices, but a great many other devices including automobiles and billions of other devices that traditionally have not had wireless connectivity. When will Cooper's Law plateau? That is: how close are we to fundamental limits of densification, where further densification no longer allows (significant) further spectrum reuse and the accompanying throughput gains? What will cause this saturation, and what can we do to prolong Cooper's Law as long as possible by optimizing the wireless network design in light of these fundamental limits? Is this even something we need to worry about?

## 3 The Bullish View on Densification

It is not actually clear that there will be a fundamental limit to densification, aside from the obvious and uninteresting scenario where the density reaches the limits of the size of the devices and no further infrastructure or devices literally can fit anywhere. We are interested in fundamental limits that assert themselves far before that scenario, and are instead due to physical limits arising from the propagation of electromagnetic signals, and a possible collapse in the Signal-to-Interference-plus-Noise ratio (SINR). Closely related but not identically, we wish to understand how the area spectral efficiency (ASE) scales with density, since the main point of densification is to increase the ASE, which is just the sum throughput normalized by the area and bandwidth.

### 3.1 Simple Example with a Square Grid

Consider a toy example of a 3 by 3 square grid with frequency reuse 1 and a BS in the middle of each cell, and the cell edges having length 2R. Thus, this network has a density of 1 base station per $4R^2$, i.e. $1/4R^2$. For example, if R = 100m for an inter-site distance of 2R = 200 meters, a fairly conservative value in a dense urban grid, then the density would be 25 BSs/km$^2$. We assume that all BSs transmit at a fixed power $P_t$ and experience *standard power law attenuation* over a distance $d$ described by $P_r(d) = P_t K_0 d^{-\alpha}$, where $P_r$ is the received power, $K_0$ a reference loss at $d = 1$, and $\alpha$ is the "path loss exponent". Consider the case where the middle base station transmits to a user at one of its four cell corners. It can be easily seen that the resulting SINR is:

$$\text{SINR} = \frac{(\sqrt{2}R)^{-\alpha}}{3(\sqrt{2}R)^{-\alpha} + 4(\sqrt{10}R)^{-\alpha} + (3\sqrt{2}R)^{-\alpha} + \sigma^2/(P_t K_0)}$$

$$= \frac{2^{-\frac{\alpha}{2}}}{3 \cdot 2^{-\frac{\alpha}{2}} + 4 \cdot 10^{-\frac{\alpha}{2}} + \cdot 18^{-\frac{\alpha}{2}} + \sigma^2 R^\alpha/(P_t K_0)}$$

which for any suitably dense scenario has negligible noise power $\sigma^2$ compared to the interference, and so the noise term can be ignored, and SINR = SIR. Clearly, the SIR is independent of R for this setup, meaning that the received signal quality does not depend on the cell size. Furthermore, this intuition holds for a square grid of any size.

### 3.2 SINR Invariance and Cell Splitting Gain

This same distance independence for SIR can be reproduced in much more general settings, and holds not only at the cell edge, but over the entire cell. The SINR is more generally a complex random variable depending on random variables such as the user's location in the cell, its distance from all the interfering base stations, and random channel effects such as fading and shadowing on all those links. Nevertheless, the SINR distribution can be derived analytically for the case of Poisson distributed base stations [AndBac11]. Furthermore, it has been recently shown that a very wide class of spatial BS distributions, including the hexagonal grid, have nearly the exact same SIR statistics as the Poisson case, with just a small fixed SIR shift (e.g.

1.5-3 dB) [GuoHaenggi15]. Somewhat incredibly, the entire SIR distribution is essentially density independent given the model discussed so far, particularly the standard path loss model.

From these results one concludes that the SINR distribution increases with the density up to the point where noise becomes negligible, after which it becomes equal to the SIR and independent of the base station density. This interesting property has been referred to as *SINR Invariance*. SINR invariance even holds in a multi-tier cellular network, where each tier of base stations (e.g. macro, micro, pico, femto) has a different transmit power and density, as long as the network is open access and users connect to the strongest signal [DhiGan12]. If SINR invariance exists (or a close approximation to it), then this is a strong motivation to densify the network, since it means that cells can be split indefinitely through the addition of new base stations, while maintaining the same SINR distribution in the network.

Since each user maintains the same SINR statistics, but shares its base station with an ever smaller number of other users, each user can seemingly achieve approximately linear growth in its achievable data rate as base stations are added. This is referred to as *cell splitting gain*. In the inevitable case that cells are split imperfectly so that more users remain on one base station than another, the gain is less than exactly proportional. However, much of that loss in cell-splitting gain can be recovered by appropriate load balancing (e.g. via biasing and interference management) [AndSin14] and by the fact that the lightly loaded base stations will not transmit as often, thus reducing the amount of interference when they are inactive.

All of this seems very encouraging, and indeed it is correct up to a point. Densification increases wireless network throughput in nearly all cases in practice to date. But can we densify indefinitely?

## 4   Revisiting Path Loss Models

SINR invariance is quite robust to many aspects of wireless network modeling, including the base station layout (as discussed above), lognormal shadowing, several common types of fading including Rayleigh and Nakagami, and the transmit power strategy. It also holds under multiple antenna enhancements, including sectorization [BlaKar16] and space division multiple access [DhiKou13]. However, it turns out that there is one key and possibly surprising modeling aspect that SINR invariance is very fragile to: the standard power law path loss model with a single path loss exponent.

The standard path loss model is nearly ubiquitous: from research to textbooks, from industry simulations to the development of standards. But it is not clear it is a good model, especially for dense wireless networks. Little attention has been paid to short-range wireless channel behavior because the radios have always assumed to operate with adequate SINR, and so the focus was on longer-range communication. Short-range wireless channels exhibit quite different behaviors than their classical cellular counterparts. In particular, the troubling effects of *path loss subduction* could disrupt densification gains.

## 4.1 Path Loss Subduction

*Path loss subduction* is the reduction of the path loss exponent in multi-breakpoint models for a radio environment as the transmitter-receiver (TR) separation distance decreases. Plots of subducted path loss as a function of TR separation distance have a slope that flattens or even reverses polarity at close range. Subducted path loss regions with an exponent less than 2 have been measured for both indoor [MacLam93] and outdoor environments [AbdAya14].

One basic physical explanation for path loss subduction can be illustrated by the two-ray propagation model in Fig. 1, which illustrates the cross-section of a wireless link with a transmitter antenna at height $h_t$ and a receiver antenna at height $h_r$, the two separated by distance $r$. The propagation in this scenario is described by the classical two-ray model, which states that the bulk of the received power is described by a line-of-sight component and a similar-magnitude ground reflection than travels a slightly longer path distance $d_i$.

There are several interesting regions of different path loss behavior in Fig. 1. Most cellular systems are traditionally assumed to operate in the *ground Fresnel region*, where the direct and reflected waves destructively interfere with one another to produce path loss closer to $\alpha=4$ rather than free space $\alpha=2$. This has always been a naïve description of realistic propagation, however, since there are likely numerous obstructions and scatterers in between the BS and mobile in these scenarios. Better, more physically correct descriptions of cellular propagation attribute excess losses to blockages from and diffraction over nearby scatterers.

The two-ray model is a better description of short-range propagation, where there are fewer intervening obstructions. This presents a problem, however, because the behavior of radio waves transitions from the ground Fresnel region to a *large-scale interference region*. In the large-scale interference region, the direct line-of-sight wave and the ground reflected waves can add either constructively or destructively. On average, this results in path loss subduction as the path loss exponent $\alpha$ returns to a value closer to free space.

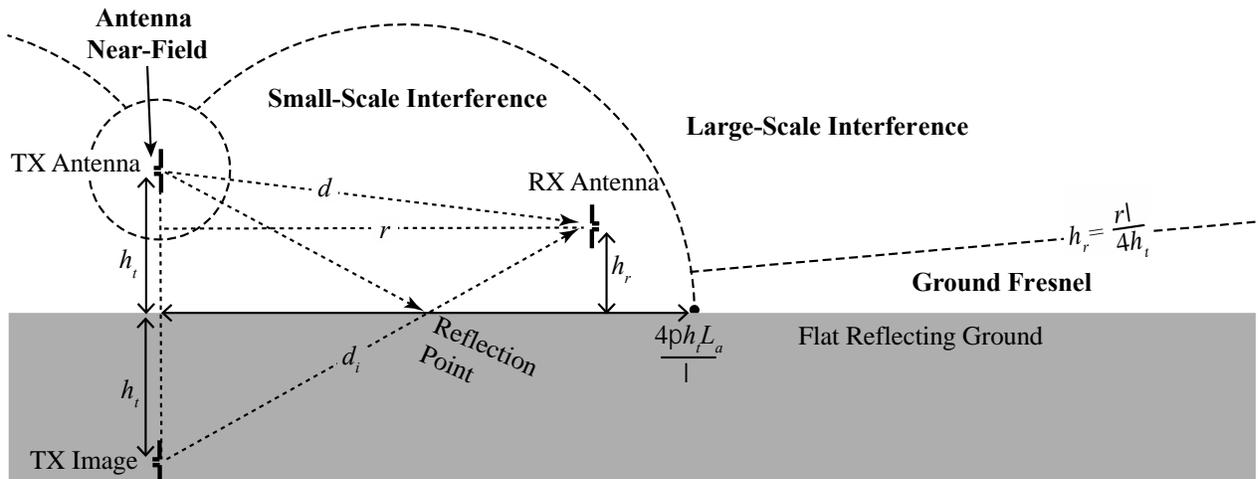

**Figure 1: Cross-section of two-ray propagation from a transmitter to a receiver, separated by a distance $r$ that may place it in one of several distinct propagation regimes.**

Even more challenging is the *small-scale interference* region, which occurs at reduced transmitter-receiver separation distances. In the small-scale interference region, the direct and reflected components result in constructive-destructive wave interference effects that fluctuate power over small-scale movements in a mobile unit (centimeters for most cellular frequencies). With even fewer possible obstructions, the total average power in this scenario will be due to the line-of-sight radio wave and the reflected radio wave (with similar magnitude). Throw in additional multipath from nearby scatterers and the path loss exponent now typically drops *below* that of free space.

The small-scale interference region forms a bubble around a transmitter and can be somewhat arbitrary in size, depending on how one defines the threshold distance of small-scale power fluctuations, $L_a$. For a given wavelength $\lambda$ (inversely proportional to frequency) and transmitter height $h_t$, this region occurs for transmit-receive separation distances less than $\frac{4\pi L_a}{\lambda}$. Table I presents several separation distances for common wireless networking frequencies and configurations for a fluctuation distance of $L_a$=20 cm. Clearly, the small-scale interference region of propagation becomes important for shorter distances. This region also becomes more important for higher frequencies, implying that mmWave systems will need to contend with similar behavior.

**Table I: Distances where the small-scale interference region ends for various wireless network frequencies and configurations, based on a threshold of significant power fluctuation over 20cm of mobile position changes.**

| TRANSMITTER TYPE | FREQUENCY $f$ | WAVELENGTH $\lambda$ | TRANSMIT HEIGHT $h_t$ | SMALL-SCALE INTERFERENCE DISTANCE |
|---|---|---|---|---|
| Cellular Macrocell | 860 MHz | 35 cm | 60 m | 432 m |
| 802.11b Access Point | 2.4 GHz | 12.5 cm | 3 m | 60 m |
| 802.11a Access Point | 5.8 GHz | 5.2 cm | 3 m | 146 m |
| LTE microcell | 700 MHz | 43 cm | 5 m | 29 m |
| future mmWave femtocell | 60 GHz | 5 mm | 2 m | 1005 m |

## 4.2 The Dual Slope Path Loss Model

A simple model which can capture the path loss subduction effect is the *dual slope path loss model*:

$$P_r(d) = \begin{cases} P_t K_o d^{-\alpha_0}, & d \leq R_c \\ P_t K_1 d^{-\alpha_1}, & d > R_c \end{cases}$$

Compared to standard path loss attenuation, this model still has power law path loss, but introduces a *corner distance* $R_c$, below which we have path loss exponent $\alpha_0$ and above which we have $\alpha_1 \geq \alpha_0$, and $K_1$ is a constant to ensure continuity between the two regions. Each path loss exponent $\alpha_i$ results in a different slope of -10$\alpha_i$ in a dB scale. Note that this model is more general than standard path loss and reverts to it for $\alpha_1 = \alpha_0$, or for $R_c = 0$ and $\alpha = \alpha_1$, or for $R_c = \infty$ and $\alpha = \alpha_0$. It also captures the classical 2-ray ground reflection model with $\alpha_0 = 2$ and $\alpha_1$

= 4, and can generally capture the breakpoint effects discussed previously stemming from path loss subduction. The dual slope model can clearly better approximate any distance dependent path loss trend, given the extra two parameters $R_c$ and $\alpha_1$ versus the standard path loss model.

In addition to better capturing path loss subduction, a multi-slope path loss model finds empirical support dating back at least until 1989 for indoor [Akerberg89] and outdoor [FeuBla94] propagation scenarios, and has been introduced recently by 3GPP to model many scenarios. These particularly include trying to differentiate between LOS and NLOS propagation, where $\alpha_0$ is the LOS path loss exponent and $\alpha_1$ is the NLOS exponent. For example, in the Urban Microcell (UMi) model $R_c$ values of 20 and 200 meters are considered as the switching point between LOS and NLOS propagation [LiuXia16]. It also can describe the blocking-based path loss models commonly used for mmWave systems [BaiHea15], in that case *$R_c$* is a random variable where blocking occurs separating the LOS and NLOS regions.

Intuitively, it might seem that cellular network performance should improve with a two-slope model, since the desired transmitter would often be the closest one, and thus experience more benign attenuation as opposed to the bulk of the interferers, who would be farther than $R_c$. And indeed, if the cell boundaries are on the order of $R_c$, the SINR is better than with a single slope model with path loss exponent $\alpha_0$ or $\alpha_1$. However if the cell radius is much larger than $R_c$, then most of the attenuation (both desired and interfering signals) is according to $\alpha_1$. Similarly if the cell size is much smaller than $R_c$ then a significant number of interferers will lie within $R_c$ – termed the *close in* region – and thus experience benign attenuation following $\alpha_0$. Therefore, the dual slope model appears useful for understanding the effect of network densification, which pushes more and more of the transmitters into a benign attenuation regime – such as line-of-sight where $\alpha_0 = 2$ might be typical – as opposed to a harsher propagation environment where $\alpha_1$ is on the order of 3 or 4, as is typically assumed.

## 5   Losing SINR Invariance: Densification with a Dual Slope Model

Our recent work [ZhaAnd15] has derived the SINR distribution for a cellular network with a multislope model, and considered how the SINR and the potential throughput as the network becomes very densely populated with base stations. We've more recently extended these results to 3 dimensions, including the so-called 3D$^+$ case where a mobile user on the ground experiences interference from base stations extending in the positive vertical direction due to dense urban environments [GupZha15]. A related scenario is the case of unmanned aerial vehicles or balloons serving as base stations, as being currently pursued by several companies including Facebook and Google, as detailed by several articles in the May 2016 issue of this magazine.

### 5.1   SINR Scaling Behavior

The key fact shown by [ZhaAnd15] is that *SINR invariance does not hold with 2 or more slopes*. Instead, the SINR at first increases as BSs are added, since the network is noise-limited at low densities. This is not surprising. However, after some finite density BS density $\lambda^*$ is reached, at which the network is no longer noise-limited, from there onwards *the SINR decreases monotonically as the density increases*.

This behavior under a dual-slope model is conceptually the opposite of SINR invariance. In fact, if the close-in path loss exponent $\alpha_0 \leq 2$ in the 2D plane or if $\alpha_0 \leq 3$ in the 3D or 3D+ case, the SIR goes to zero for all users in the system as the density gets very large. We refer to these values as *critical path loss exponents*. More formally, the probability of a user achieving a given non-zero SIR goes to zero if $\alpha_0$ is below the critical path loss exponent; and this probability goes to zero particularly fast for moderate-to-high SIRs. We can observe all these trends in Figure 2 for the 2D case, where we observe empirically that this transition occurs at a density of about 10-20 BSs/km². Up until this density the SINR monotonically increases as the SNR improves, and after it the SINR monotonically decreases as the network becomes ever more interference-limited.

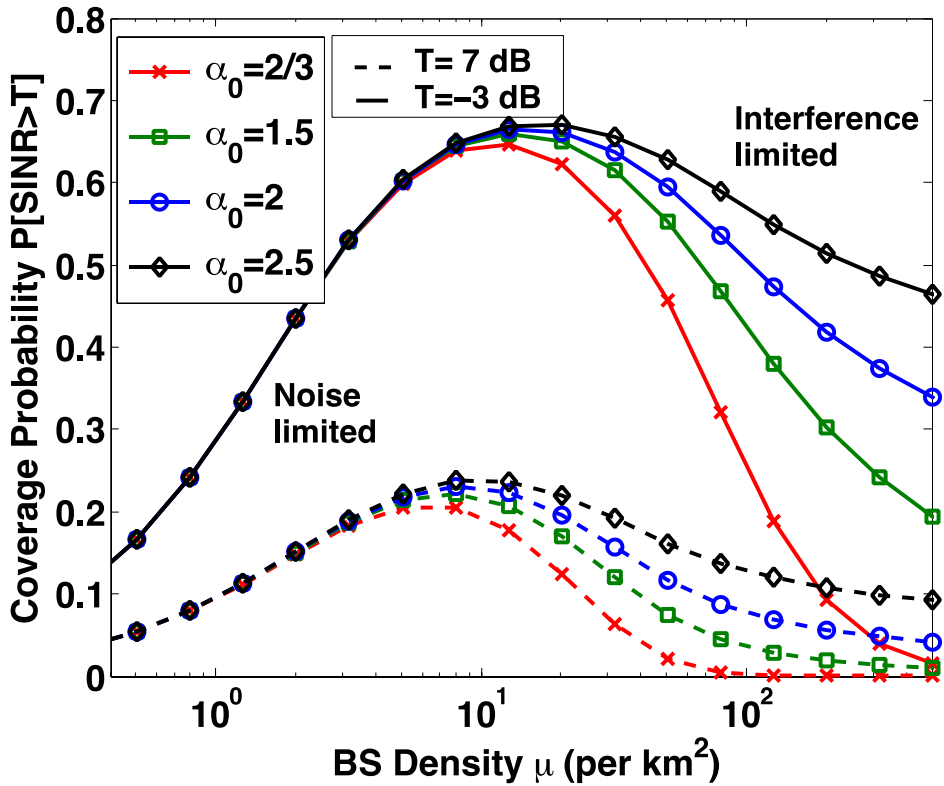

**Figure 2: Probability of coverage vs. BS density (Poisson distributed, average per km²) in the 2D plane for SINR threshold T = 0.5 (-3dB) and 5 (7 dB), close-in path loss exponent $\alpha_0$, and with $R_c$ = 100 m and $\alpha_1$ = 4. The noise is set in each case such that the SNR=20 dB at $R_c$. There is Rayleigh fading, but no shadowing or sectoring.**

These trends are visible even when $\alpha_0$ is above the critical path loss exponent; just in that case the SINR monotonically decreases but never reaches zero. Further, these critical path loss exponents are operationally relevant, since in nearly all cases $\alpha_0 \leq 3$, and there are also many scenarios where $\alpha_0 < 2$ due to path loss subduction, as already discussed. Since these ultra-dense scaling results only ultimately depend on $\alpha_0$, they apply equally to a multi-slope model with any number of path loss exponents (such as [Akerberg88], which has 4). It is also clear that 10-20 BSs/km$^2$ is not even particularly dense for urban deployments, so these trends are likely already visible in real systems. These results are based on mathematical analysis, but interestingly, a simulation study in 2014 by Ericsson researchers found that around this exact same density, throughput began to increase in HSPA cellular networks [WuBut14].

## 5.2 Throughput Scaling Behavior

Even if the SINR trends to zero for high densities, the sheer number of base stations may still allow for the overall throughput in the network to increase. We define the *potential throughput* as the maximum throughput per unit area (or volume), which is achieved when all BSs are transmitting to an active user. It is shown in [ZhaAnd15] for the 2D plane that although for $\alpha_0 > 2$ the potential throughput growth is still linear (as in the standard path loss case), for $1 < \alpha_0 < 2$ the growth slows to be sublinear with density, specifically the exponent is $2-2/\alpha_0$. This result implies a significantly diminished return on investment for cellular providers that increase network density beyond a certain threshold.

Only for the unlikely case that $\alpha_0 < 1$ does the potential throughput also go to zero. However, for the 3D or 3D+ case, the critical path loss exponents increase by 50%, and so potential throughput goes to zero for $\alpha_0 < 1.5$, which is perhaps improbable in most environments but not completely impossible.

To get a better feel for when these theoretically derived behaviors might assert themselves in very dense wireless networks, we consider Figure 3 which shows the potential throughput as the BS density increases. For $\alpha_0 < 1$ or $\alpha_0 < 1.5$, for 2D and 3D respectively, the potential throughput decreases after a certain threshold of density, which we term the *critical density* $\mu_c$. Formally, this is where the derivative of the potential throughput is zero. Furthermore, we observe that the potential throughput nearly saturates for moderate SINR targets (such as T = 7 dB) even for $\alpha_0 = 2$ in the 3D case.

We also define the *normalized critical density* as the average number of BSs in the close-in region, i.e. in a ball of radius $R_c$, when the potential throughput peaks and then decreases. This normalized critical density is $\pi R_c^2 \mu_0$ for 2D and $4\pi/3 \, R_c^3 \, \mu_0$ for 3D. In Table II, we tabulate values of the critical densities and observe that not very many interferers need to be in the close-in region, on average, for these troubling behaviors to become visible.

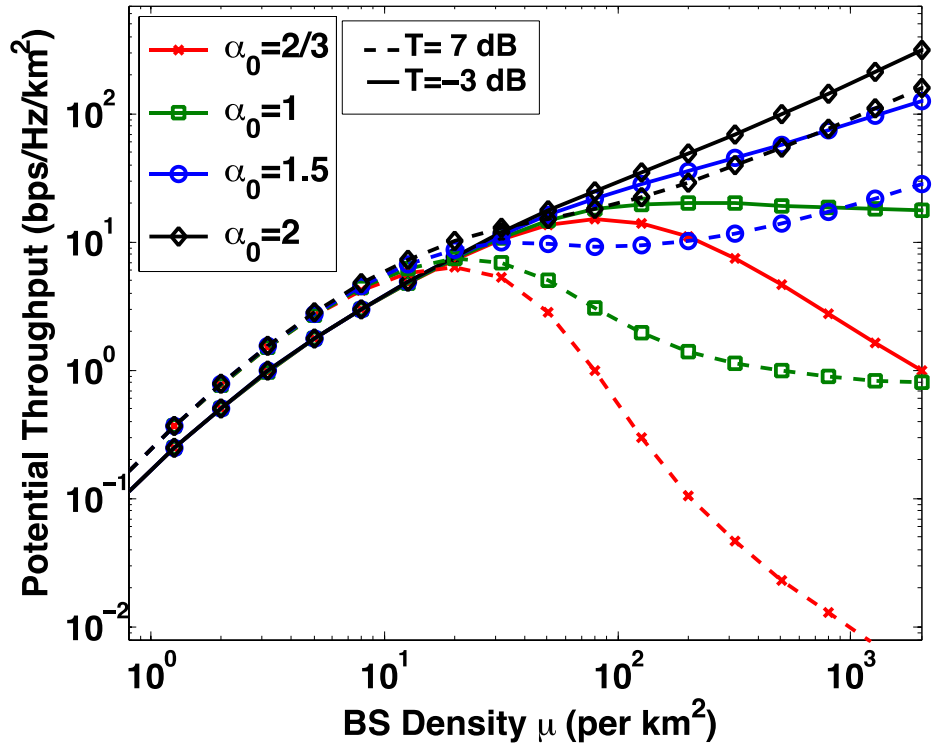

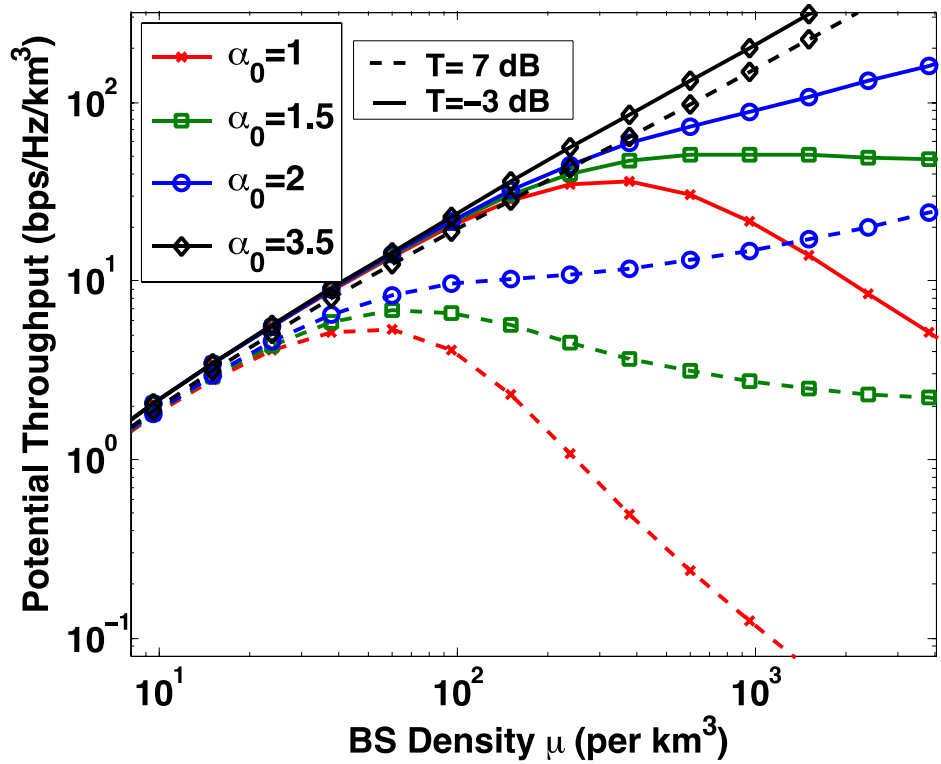

Figure 3: Potential throughput vs. BS density for 2D (top) and 3D (bottom) network topologies. Parameters are the same as for Fig. 2.

**Table II: Normalized critical BS densities (number of BSs inside $R_c$) where potential throughput begins to decrease.**

| Parameters | | | | Normalized critical BS density | |
|---|---|---|---|---|---|
| dim | $R_c$ | $\alpha_0$ | $\alpha_0$ | SINR T=0.5 (-3 dB) | SINR T=5 (7 dB) |
| 2D | 20m | 2/3 | 4 | 2.67 | 0.58 |
| 2D | 20m | 1 | 4 | 6.14 | 0.70 |
| 2D | 100m | 2/3 | 4 | 1.35 | 0.21 |
| 2D | 100m | 1 | 4 | 3.30 | 0.29 |
| 3D | 20m | 1 | 4 | 1.35 | 0.21 |
| 3D | 20m | 1.5 | 4 | 3.30 | 0.29 |
| 3D | 100m | 1 | 4 | 1.35 | 0.21 |
| 3D | 100m | 1.5 | 4 | 3.30 | 0.29 |

## 5.3 Takeaways and Caveats

Although these results are predicated on an improved but still idealized propagation model, we believe they do raise a number of interesting and pressing questions on the viability of continuing to enhance wireless network data rates primarily through network densification. However, we do wish to caution readers regarding several caveats pertaining to these results. These caveats are:

1. The base stations are "fully loaded", that is they always have an active user desiring data and so they transmit all the time. If the user density is fixed, then as the BS density grows, not all BS's will have an active user, and thus will not contribute interference.
2. All users in a given range have the same path loss exponent, which is an unrealistically homogeneous assumption on the propagation environment. Further, a user will be biased towards connecting with the base station with the most benign propagation, which means that other base stations, statistically speaking, will have their interference attenuated more.
3. Neither shadowing nor blocking have been explicitly considered. Similar to the 2[nd] caveat, the desired link would usually have benign shadowing or blocking while the nearby interfering links would have slightly worse shadowing on average, and cumulatively the shadowing and blocking will attenuate interference more than the desired signal, thus increasing SINR [BaiHea15].
4. We do not consider any attempts at interference management or suppression, which apparently will take on renewed importance in ultra-dense networks given the increased interference. Thus, these are fundamental limits only on the preprocessing SINR, since the post-processing SINR and thus the throughput can be improved along the lines of information-theoretically optimal transmit and receive techniques which mitigate interference.

## 6  Conclusions and Future Challenges

The initial results in this paper indicate that in many scenarios, electromagnetic fundamentals coupled with network level analysis indicate troubling behaviors in ultradense networks. In particular, we observe that SINR invariance is almost sure to be lost at high densities, and that instead the SINR will monotonically decrease as the network is densified past a certain point. It can even decrease to zero, if the close-in path loss exponent is below the critical exponent (2 and 3 for 2D and 3D, respectively). Similar trends are observed for the throughput, although it is more robust to densification than SINR.

Although we have painted a somewhat bleak picture of what can happen when the close-in path loss exponents are on the order of free space propagation or less, there may be many technical approaches that push out the effective critical density. Since the cause of the observed behaviors is strong aggregate interference, the obvious approaches are those that suppress interference. Although the myriad interference suppression techniques are by now quite well researched, it is worth revisiting the topic in the context of ultra-densification with appropriate propagation models. The gains of techniques such as coordinated multipoint (CoMP), interference cancellation, and resource blanking can be expected to increase in ultradense networks, perhaps significantly. Furthermore, directional transmissions may also be more robust, as are envisioned for massive MIMO and millimeter wave systems. The main point of this paper is that wireless network researchers and engineers should be aware of these rapidly approaching limits, and we should begin developing communication protocols customized for dense networks, to push these fundamental limits of densification out as far as possible.

## Acknowledgements

The authors greatly appreciate the considerable feedback and inputs they have obtained on these ideas from many people, including the system engineering teams at Huawei (Plano, TX and Paris, France), Samsung's Dallas Technology Lab, and Nokia Networks (Arlington Heights, IL). Particular thanks to Anthony Soong of Huawei for originally suggesting that we conduct analysis with a 2-slope path loss model.